\documentclass[a4paper,11pt,twocolumn]{article}

\usepackage{graphicx}
\usepackage[a4paper,lmargin=1.6cm,rmargin=1.6cm,tmargin=1.5cm,bmargin=2.7cm]{geometry}
\usepackage{authblk}
\usepackage{titlesec}
\usepackage{url}
\usepackage{caption}

\newcommand{\citep}[1]{\cite{#1}}
\newcommand{\citet}[1]{\cite{#1}}
\newcommand{\zeroindexLarge}{0}
\newcommand{\zeroindexSmall}{{\rm o}}
\newcommand{\Vzero}{V_\zeroindexLarge}
\newcommand{\nzero}{n_\zeroindexSmall}
\newcommand{\vzero}{v_\zeroindexSmall}
\newcommand{\epsilonzero}{\epsilon_\zeroindexSmall}

\title{\Large{\textbf{COULOMB DRAG DEVICES: ELECTRIC SOLAR WIND SAIL PROPULSION AND IONOSPHERIC DEORBITING}}}

\titleformat{\section}{\center\large\scshape}{\thesection}{1em}{}
\titleformat{\subsection}{\itshape}{\thesubsection}{1em}{}
\author{Pekka Janhunen}

\affil{Finnish Meteorological Institute, POB--503, FI--00101, Helsinki,
  Finland}
\date{}

\begin{document}

\maketitle

\begin{abstract}
A charged tether or wire experiences Coulomb drag when inserted into
flowing plasma. In the solar wind the Coulomb drag can be utilised as
efficient propellantless interplanetary propulsion as the electric
solar wind sail (electric sail, E-sail). In low Earth orbit (LEO) the
same plasma physical effect can be utilised for efficient low-thrust
deorbiting of space debris objects (the plasma brake). The E-sail is
rotationally stabilised while the deorbiting Coulomb drag devices can
be stabilised either by spinning or by Earth's gravity gradient.

According to numerical estimates, Coulomb drag devices have very
promising performance figures, both for interplanetary propulsion and
for deorbiting in LEO. Much of the technology is common to both
applications. E-sail technology development was carried out in ESAIL
FP7 project (2011-2013) which achieved TRL 4-5 for key hardware
components that can enable 1 N class interplanetary E-sail weighing
less than 200 kg. The thrust of the E-sail scales as inverse solar
distance and its power consumption (nominally 700 W/N at 1 au) scales
as the inverse distance squared. As part of the ESAIL project, a
continuous 1 km sample of E-sail tether was produced by an automatic
and scalable ``tether factory''. The manufacturing method uses
ultrasonic wire to wire bonding which was developed from ordinary wire
to plate bonding for the E-sail purpose. Also a ``Remote Unit'' device
which takes care of deployment and spin rate control was prototyped
and successfully environmentally tested. Our Remote Unit prototype is
operable in the solar distance range of 0.9-4 au.

The 1-U CubeSat ESTCube-1 was launched in May 2013 and it will
try to measure the Coulomb drag acting on a 10 m long tether in
LEO when charged to 500 V positive or negative. A
more advanced version of the experiment with 100 m tether is
under preparation and will be launched in 2015 with the Aalto-1
3-U CubeSat to polar LEO.
\end{abstract}

\section{Introduction}

\begin{figure}
\includegraphics[width=8.5cm]{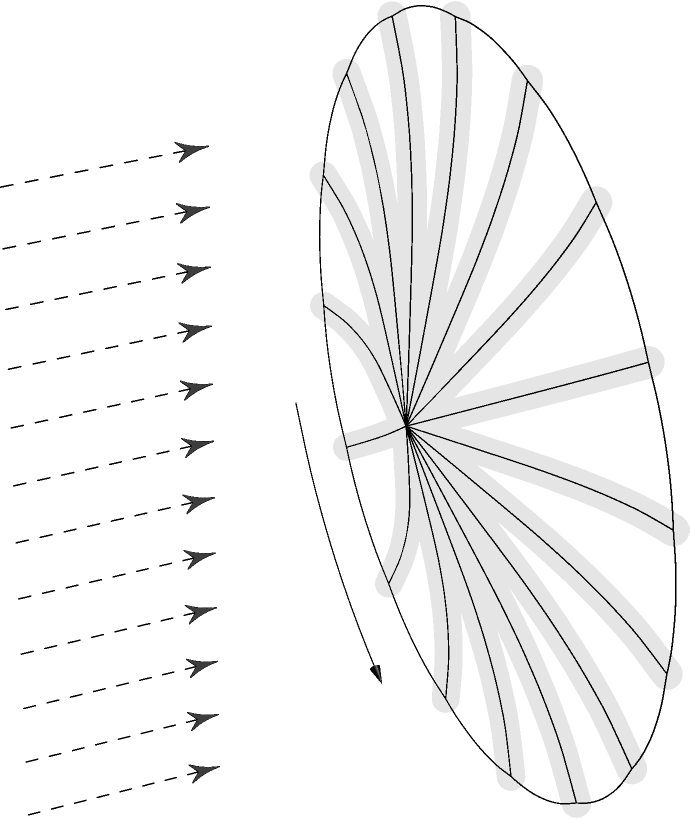}
\caption{
Centrifugally stabilised E-sail with auxiliary tethers \citep{paper9}. The auxiliary tethers connect the
tips of the main tether and prevent the main tethers from
colliding with each other despite solar wind variations. A ``Remote
Unit'' (Fig.~\ref{fig:RUpict}) is located at the tip of each main tether (not visible in this
scale). The Remote Unit hosts the auxiliary tether reels and a
thruster (e.g.~cold gas, ionic liquid FEEP or photonic blade) to
initiate the spin during deployment and to manage the spinrate during
flight if needed.
}
\label{fig:Esail3D}
\end{figure}

The concept of solar wind electric sail (electric sail, E-sail,
Fig.~\ref{fig:Esail3D}) was proposed as a device which harnesses the
momentum flux of the natural solar wind \citep{paper1,paper2}. The
first attempt to predict the thrust per unit tether length was based
on electrostatic particle-in-cell (PIC) simulations \cite{paper2}. The
initial thrust estimate was later corrected upward \citep{paper6}.

\begin{figure}
\includegraphics[width=8.5cm]{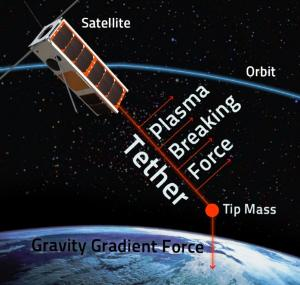}
\caption{
Negatively biased gravity-stabilised Coulomb drag plasma brake in LEO
for satellite deorbiting \citep{KestilaEtAl2013}.
}
\label{fig:PlasmaBrake}
\end{figure}

It was realised that not only a positive, but also a negative tether
would produce a Coulomb drag effect \citep{paper3}. As a new
application for the technology, it was proposed that a single
gravity-stabilised negatively biased tether could be used for
satellite and space debris deorbiting in low Earth orbit (LEO)
\cite{plasmabrake} (Fig.~\ref{fig:PlasmaBrake}).

The E-sail has potentially revolutionary performance level in
comparison to other propulsion systems \citep{paper9}. The tether
weighs only 10 grams per kilometre and produces a thrust of $\sim 0.5$
mN/km at 1 au distance. The E-sail thrust scales as proportional to
$1/r$ where $r$ is the solar distance. The reason is that while the
solar wind dynamic pressure decays as $\sim 1/r^2$, the plasma Debye
length (by which the electric field penetration distance and hence the
virtual sail size scales) varies as $\sim r$, thus giving an overall
$1/r$ dependence for the thrust. For example, hundred 20 km long
tethers would weigh 20 kg and they would produce 1 N thrust at 1 au
which gives a 30 km/s velocity change pear year for a 1000 kg
spacecraft.

E-sail thrust magnitude can be easily controlled between zero and a
maximum value by changing the voltage of the tethers. The tether
voltage is maintained by continuously operating an electron gun which
pumps out negative charge from the system, hence tether voltage can be
actuated easily by changing the current and voltage of the electron
gun beam. The power consumption of the electron gun is moderate (700 W
nominally at 1 au for large 1 N sail) and it scales as $1/r^2$,
i.e.~in the same way as the illumination power of solar panels. The
power consumption stems from the electron current gathered from the
surrounding solar wind plasma by the positively charged tethers which
can be estimated by so-called orbital motion limited (OML) cylindrical
Langmuir probe theory \cite{paper2},
\begin{equation}
\frac{d I}{d z} = e \nzero \sqrt{\frac{2 e \Vzero}{m_e}} \left(2 r_w^{\rm base} +
4.5\times r_w^{\rm loop}\right).
\label{eq:dIdz}
\end{equation}
Here $e$ is the electron charge, $\nzero$ is the solar wind plasma
density, $\Vzero$ is the tether bias voltage, $m_e$ is electron mass,
$r_w^{\rm base} = 25\,\mu$m is the base wire radius of the
micrometeoroid-resistant 4--wire Heytether
\cite{SeppanenEtAl2011,SeppanenEtAl2013}, $r_w^{\rm loop} =
12.5\,\mu$m is the loop wire radius
\cite{SeppanenEtAl2011,SeppanenEtAl2013} and $dI/dz$ is the gathered
current per unit length of the tether.  Clearly, to minimise power
consumption, the tether outer surface area should be made as small as
possible while keeping the tethers micrometeoroid-tolerant and their
resistance low enough to avoid significant ohmic power losses. The
ultrasonically bonded 4-wire aluminium Heytether satisfies these
requirements \citep{SeppanenEtAl2011,SeppanenEtAl2013}.

The thrust direction of the E-sail can be changed by inclining the
sail with respect to the solar wind flow. Tilting of the spinplane of
the tethers can be actuated by modulating the voltages of the tethers
differentially so as to produce net torque which turns the sail. Such
tilting manoeuvres typically take a few hours. Even without applying
closed loop control for sail orientation during the journey, it was
shown by Monte Carlo calculations with real measured solar wind data
that a planet such as Mars can be reached by using a simple controller
which increases (decreases) the tether voltage when the spacecraft is
behind (ahead of) schedule on its planned trajectory towards the
target \citep{paper5}. Thus, although the E-sail's thrust source is
the highly variable and basically unpredictable solar wind, its
navigability can be made comparable to other propulsion systems. In
addition, the E-sail's thrust magnitude and thrust direction can be
varied independently of each other which is a benefit in comparison
e.g.~to the photon sail where they typically vary in unison.

In missions which spiral inward or outward in the solar system there
is a secular tendency due to orbital Coriolis force for the spin rate
to decrease or increase, respectively \citep{paper14}. Because the
E-sail effect itself in general cannot produce propulsion within the
spin plane, the Coriolis effect must be counteracted by auxiliary
propulsion system on the Remote Units installed on tips of the main
tethers \citep{paper9}. To this end, in the ESAIL FP7 project, a
prototype Remote Unit was developed with two alternative propulsion
systems (cold gas and ionic liquid FEEP) \citep{ESAIL-D41.2}. Photonic
blade auxiliary propulsion devices have also been considered for this
task \citep{paper16}.

The E-sail has potentially revolutionary level of performance and
hence it has a large number of applications in solar system missions
\citep{MengaliEtAl2008,MengaliAndQuarta2009,QuartaAndMengali2010a,QuartaAndMengali2010b,QuartaEtAl2011,MerikallioAndJanhunen2010,QuartaEtAl2012,UranusPaper}. The
various E-sail solar system applications are discussed and summarised
by \citet{PEASpaper} and a readable yet rigorous presentation of the
mission possibilities from the orbital calculations perspective is
available \citep{ESAIL-D62.1}.

\begin{figure}
\includegraphics[width=8.5cm]{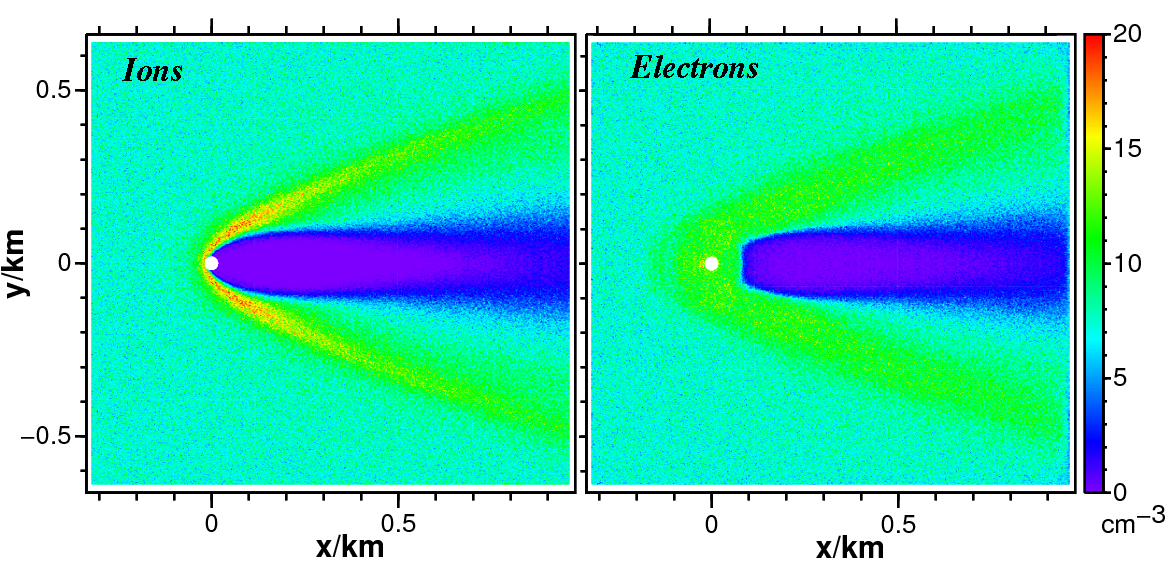}
\caption{
PIC simulation of positively charged E-sail tether interacting with streaming
solar wind plasma \citep{ASTRONUM2011}. Solar wind (density 7.3 cm$^{-3}$, speed
400 km/s) arrives from the left and interacts with +5.6 kV charged
tether at (white dot).
}
\label{fig:PositiveTether}
\end{figure}

\section{Physics of Coulomb drag}

\begin{figure}[h]
\begin{center}
\includegraphics[width=8.3cm]{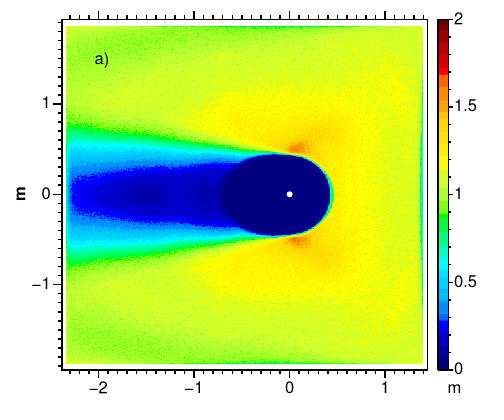}
\includegraphics[width=8.3cm]{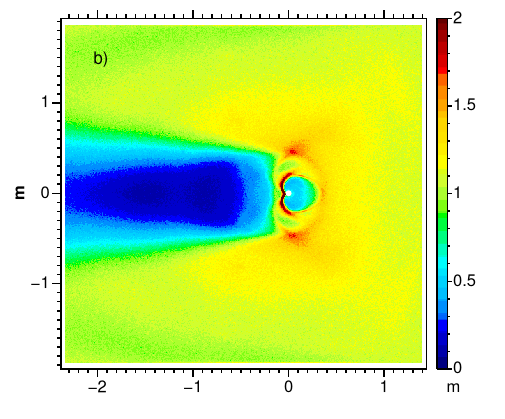}
\end{center}
\caption{Electron density (a) and ion density (b) normalised to plasma
  stream density in the final state ($t=2.43$ ms) in LEO conditions \cite{paper24}. Flow arrives from the right.}
\label{fig:NegativeTether}
\end{figure}

When plasma streams past a charged thin tether, the tether's electric
field penetrates some distance into the plasma and deflects the
charged particles of the stream. Because electrons are lightweight,
the momentum flux carried by them is negligible so it is enough to
consider the deflection of ions. Both positively and negatively biased
tethers cause ion deflection and hence Coulomb drag. A positive tether
deflects positively charged ions by repelling them
(Fig.~\ref{fig:PositiveTether}). A negative tether deflects ions by
attracting them so that their paths cross behind the tether
(Fig.~\ref{fig:NegativeTether}).

\subsection{Positively biased tether}

A positively biased tether repels stream ions and attracts
electrons. When the potential is turned on, a population of
trapped electrons gets formed \cite{paper2}. In most of the literature
concerning biased tethers, it is implicitly or explicitly assumed that
trapped electrons are not present in the asymptotic state. In a
multi-tether starfish-shaped E-sail geometry (Fig.~\ref{fig:Esail3D}),
trapped electron orbits are chaotised whenever the electron visits the
central ``hub'' which is the spacecraft, and chaotised electrons have
a small nonzero probability of getting injected into an orbit which
takes them to collision course with a tether wire so that trapped
electrons are removed by this mechanism in few minute timescale in
nominal 1 au solar wind \cite{paper6}. It might be that other
processes such as plasma waves occur in nature which speed up the
process. By PIC simulations alone it is not easy to predict how much
if any trapped electrons are present in the final state.

The E-sail thrust per tether length $dF/dz$ is given by
\begin{equation}
\frac{dF}{dz} = K P_{\rm dyn} r_s
\label{eq:dFdz}
\end{equation}
where $K$ is a numerical coefficient of order unity, $P_{\rm dyn} =
\rho v^2$ is the dynamic pressure of the plasma flow and $r_s$ is the
radius of the electron sheath (the penetration distance of the
electric field into the plasma). The quantity $r_s$ can be inferred
from recent laboratory measurements of Siguier et
al.~\cite{SiguierEtAl2013} where Ar$^{+}$ plasma (ion mass $m_i=40$ u) of
density $\nzero=2.4\cdot 10^{11}$ m$^{-3}$ accelerated to 20 eV bulk flow
energy (hence speed $v=9.8$ km/s) was used and let to interact with $r_w=2.5$ mm
radius metal tether biased to $\Vzero=100$ V and 400 V in two experiments. At
$\Vzero=100$ V the sheath radius as visually determined from their Figure
7 is $r_s=12$ cm and at 400 V it is $r_s=28$ cm (from their Figure 8). For estimating
the corresponding $dF/dz$, let us assume $K=2$ in the above formula
[Eq.~(\ref{eq:dFdz})]. This corresponds to assuming that ions incident
on the sheath are on average deflected by 90$^{\circ}$ (notice that
the size of the virtual obstacle made by the sheath is twice its
radius). We think that this is a reasonable first estimate since ions
arriving head-on towards the tether are reflected backwards while ions
arriving near the boundaries of the sheath are probably deflected by
less than 90$^{\circ}$. In their experiment $P_{\rm dyn} = 1.54$ $\mu$Pa so
Eq.~\ref{eq:dFdz} gives $dF/dz=370$ nN/m and $dF/dz=860$ nN/m for
$\Vzero$ equal to 100 V and 400 V, respectively.

Let us compare these experimentally inferred values with theoretical
estimates. A simple theoretical estimate for the sheath radius is the
effective Debye length
\begin{equation}
\lambda_D^{\rm eff} = \sqrt{\frac{\epsilon_0
    \left(\Vzero-V_1\right)}{e \nzero}}
\label{eq:lambdaDeff}
\end{equation}
where $V_1=(1/2)m_i v^2/e$ is the stream ion bulk flow energy. The
expression (\ref{eq:lambdaDeff}) for the effective Debye length is
obtained from the usual formula for ordinary electron Debye length by
replacing the electron temperature by the tether voltage. We also
subtract the bulk energy term $V_1$ to model the fact that if the
tether voltage is lower than the bulk energy, it can no longer reflect
back or stop ions but only weakly deflects them even if they arrive
with zero boost parameter; the subtraction of $V_1$ however has only
modest impact to our results. If one takes $r_s$ to be equal to
$\lambda_D^{\rm eff}$ in Eq.~(\ref{eq:dFdz}), one obtains $dF/dz$
equal to 420 nN/m and 910 nN/m for $\Vzero$ equal to 100 V and 400 V,
respectively.

Theoretical E-sail thrust formulas of \cite{paper9} contain the average
electron density $n_e$ inside the sheath as a free parameter, the
choice $n_e=0$ giving the largest E-sail thrust. Assuming $n_e=0$ and
applying the formulas for the experimental parameters of Siguier et
al.~\citep{SiguierEtAl2013}, one obtains 220 nN/m and 740 nN/m thrust
per length for $\Vzero$ equal to 100 V and 400 V, respectively.

We summarise the experimental and theoretical results in Table
\ref{tab:numeric-comparison}.

\begin{table}
\caption{Comparison of experimental and theoretical E-sail thrust per length in LEO-like conditions.}
\label{tab:numeric-comparison}
\begin{tabular}{lll}
\hline
                                               & $\Vzero$=100 V & $\Vzero$=400 V  \\
\hline
Siguier et al.~\citep{SiguierEtAl2013}         & 370 nN/m       & 860 nN/m        \\
$\lambda_D^{\rm eff}$, Eq.~\ref{eq:lambdaDeff} & 420 nN/m       & 910 nN/m        \\
Theory of \citep{paper9} /w $n_e$=0            & 220 nN/m       & 740 nN/m        \\
\hline
\end{tabular}
\end{table}

We conclude from Table \ref{tab:numeric-comparison} that experimental
results of \citep{SiguierEtAl2013} are consistent with the assumption
of no trapped electrons, i.e.~maximal E-sail thrust.

\subsection{Negatively biased tether}

\label{subsect:negbias}

In the negative polarity case, electrons are simply repelled by the
tether (Fig.~\ref{fig:NegativeTether}) and hence the physics of
electrons is simple. We believe that PIC simulations
therefore have a good chance of predicting the thrust correctly in the negative polarity case. Using a new
supercomputer, a comprehensive set of negative polarity PIC
simulations for LEO-like parameters was recently conducted
\citep{paper24} and it was found that the following formula gives a
good fit to the PIC simulations:
\begin{equation}
\frac{dF}{dz} =
3.864 \times P_{\rm dyn} \sqrt{\frac{\epsilonzero \tilde{V}}{e \nzero}}
\exp\left(-V_i/\tilde{V}\right).
\label{eq:dFdzneg}
\end{equation}
Here $\vzero$ is the ionospheric plasma ram flow speed relative to spacecraft (assumed to be perpendicular to
the tether or else $\vzero$ denotes only the perpendicular component), $P_{\rm dyn} = m_i \nzero \vzero^2$ is the flow dynamic pressure,
$m_i$ is the ion mass (typically the plasma is singly ionised atomic
oxygen so that $m_i\approx 16$ u),
\begin{equation}
\tilde{V} = \frac{\vert\Vzero\vert}{\ln(\lambda_D^{\rm eff}/r_w^{*})},
\end{equation}
$r_w^{*}$ is the tether's effective electric radius (Appendix
  A of \citep{paper2}), $\lambda_D^{\rm eff}=\sqrt{\epsilonzero \vert\Vzero\vert/(e \nzero)}$
is the effective Debye length and $V_i=(1/2)m_i \vzero^2/e$ is the
bulk ion flow energy in voltage units. The effective electric radius
is approximately given by $r_w^{*} = \sqrt{b r_w} \approx 1$ mm, where $r_w$ is the
tether wire radius, typically 12.5-25 $\mu$m, and $b$ is the tether
width, typically 2 cm (a rough value of $b$ is sufficient to
  know because $r_w^*$ enters into
  Eq.~(\ref{eq:dFdzneg}) only logarithmically).

Thus, although experimental confirmation is needed, there is good
reason to believe that Eq.~\ref{eq:dFdzneg} describes LEO plasma brake
thrust well. The only exception is that if the geomagnetic field
is predominantly oriented along the tether, the interaction becomes
turbulent and the thrust is moderately reduced \cite{paper24}. The
reduction grows with increasing voltage: it is 17\% at $-320$ V bias
voltage and 27\% at $-760$ V. For a vertical gravity-stabilised plasma
brake tether in polar orbit, efficiency reduction with respect to
Eq.~(\ref{eq:dFdzneg}) is thus expected at high latitudes.

We emphasise that Eq.~(\ref{eq:dFdzneg}) has thus far only been
verified with simulations in LEO plasma environment conditions. If
negative polarity Coulomb drag devices would become relevant in the
future also in other plasma conditions, the applicability of
Eq.~(\ref{eq:dFdzneg}) should be considered carefully on a case by
case basis.

\begin{figure}
\includegraphics[width=8.5cm]{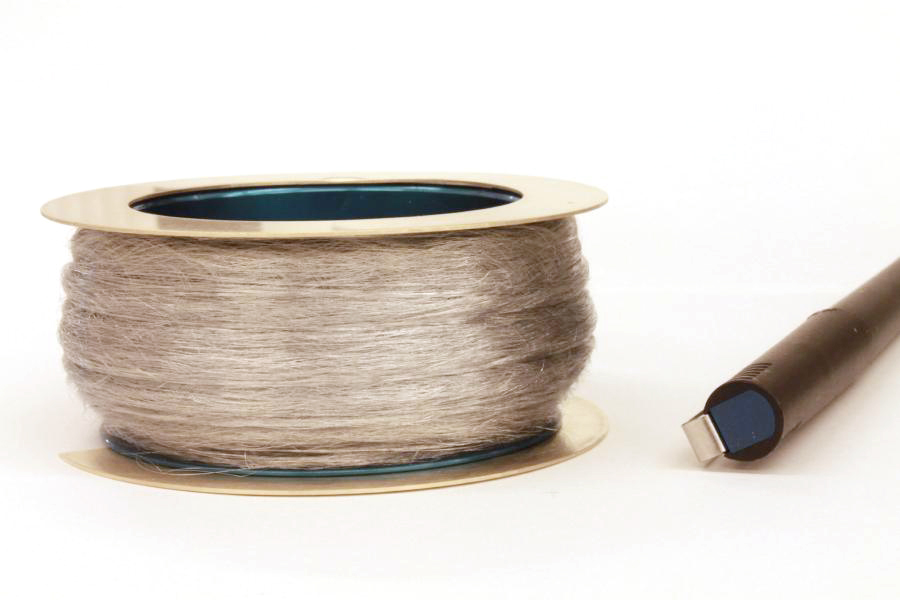}
\caption{
1 km tether (mass 11 grams) on its $\emptyset=5$ cm reel \citep{SeppanenEtAl2013}.
}
\label{fig:1kmreel}
\end{figure}

\begin{figure}
\includegraphics[width=8.5cm]{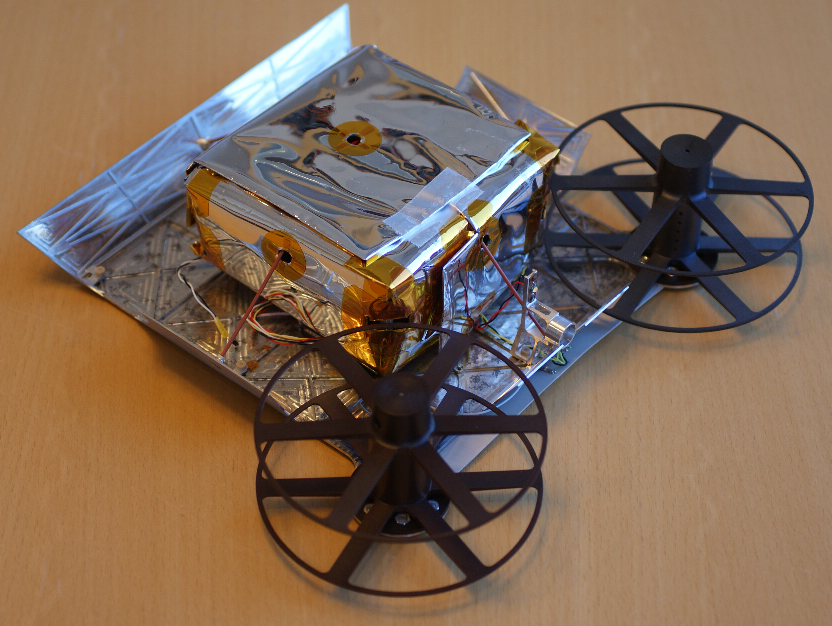}
\caption{
ESAIL project Remote Unit prototype \citep{ESAIL-D41.2}.
}
\label{fig:RUpict}
\end{figure}

\begin{figure}
\includegraphics[width=8.5cm]{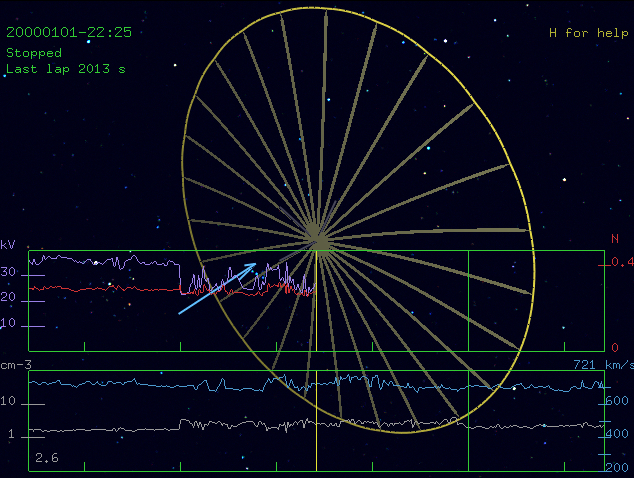}
\caption{
Screenshot from E-sail dynamical simulator \citep{ESAIL-D51.1}.
}
\label{fig:dynsim-screen}
\end{figure}

\section{E-sail development}

The goal of ESAIL FP7 project (2011-2013) was to develop prototypes of
key components of a 1 N E-sail at technical readiness level (TRL)
4-5 \citep{ESAIL-docs}. The main achievements were the following:
\begin{itemize}
\item Demonstrated manufacturing of 1 km continuous 4-wire E-sail tether (Fig.~\ref{fig:1kmreel}, \citep{SeppanenEtAl2013}).
\item Demonstrated successful reel-out of 100 m tether at low (0.25 gram) tension, also after reel was 
shaken in launch vibration test facility \citep{ESAIL-D32.4}.
\item Built and tested low-mass prototype Remote Unit with 0.9-4 au operational range (Fig.~\ref{fig:RUpict}, \citep{ESAIL-D41.2,ESAIL-D41.4}).
\item Proposed and prototyped solution for auxiliary tether (perforated kapton tape, \citep{ESAIL-D24.1}).
\item E-sail ``flight simulator'' software
  (Fig.~\ref{fig:dynsim-screen}, \citep{ESAIL-D51.1,ESAIL-D51.2}), E-sail failure mode and recovery
  strategy analysis \citep{ESAIL-D53.1}, E-sail mass budget estimation
  \citep{proj7a}, applications \citep{ESAIL-D62.1}.
\end{itemize}

These achievements suggest that a large-scale (up to 1 N) E-sail is
feasible to build. While a lot of work certainly remains to be done
before a large-scale E-sail becomes a reality, no unsolved technical
problems are currently known.

\section{Plasma brake development}

Recall from section \ref{subsect:negbias} that for LEO parameters,
Eq.~\ref{eq:dFdzneg} is a good fit to PIC simulations (except for
modest thrust reduction due to turbulence when the dominant component
of the geomagnetic field is along the tether) which in turn are
expectedly good models of reality in case of negative tether voltage.

One noteworthy fact is that LEO plasma brake thrust according to
Eq.~\ref{eq:dFdzneg} is proportional (through linear dependence on
$P_{\rm dyn}$) to the ion mass $m_i$. Thus, plasma brake thrust is 16
times larger in pure oxygen O$^{\circ}$ plasma than in pure proton
plasma.

For example, at $\Vzero=-1$ kV, $\nzero=3\cdot 10^{10}$ m$^{-3}$,
$v=7.5$ km/s and $m_i=16$ u, Eq.~\ref{eq:dFdzneg} gives 85 nN/m
thrust. In the negative bias case, usable tether voltage is limited by
onset of electron field emission. We think that above 1-2 kV, field
emission might start to become an issue.

\begin{table}
\caption{Predicted plasma brake thrust (nN/m) for solar min/solar max
  conditions.}
\label{tab:plasmabrakethrust}
\begin{tabular}{llll}
\hline
Altitude & MLT 12-00 & MLT 06-18  & Average \\
\hline
700 km & 47/157    & 42/140     & 44/149  \\
800 km & 33/117    & 30/108     & 32/112  \\
900 km & 25/88     & 22/84      & 23/86   \\
\hline
\end{tabular}
\end{table}

Table \ref{tab:plasmabrakethrust} gives plasma brake thrust based on
Eq.~\ref{eq:dFdzneg}, assuming $\Vzero=-1$ kV, $v=7.5$ km/s and using
plasma density and chemical composition taken from the IRI-2012
ionospheric model, for noon-midnight (mean local time MLT 12-00) and
dawn-dusk (MLT 06-18) polar orbits and for solar minimum and maximum
ionospheric conditions. We see from Table \ref{tab:plasmabrakethrust}
that the dependence on solar cycle is relatively significant, about
factor 3.5. The solar cycle dependence is due to increased plasma
density and increased oxygen abundance during solar maximum
conditions. There is obviously also an altitude dependence. Below 700
km the thrust would continue to increase until $\sim 400-500$ km,
provided that the hardware is designed to take advantage of it. The
dependence on MLT is weak.

As a numeric example, consider a 10 km long plasma brake tether which
starts bringing down a debris object of 200 kg mass from 800 km
circular orbit in an MLT which is average between dawn-dusk and
noon-midnight. The required $\Delta v$ from 800 km to 700 km is 53.5
m/s and from 700 km to 400 km 165 m/s. During solar minimum,
deorbiting from 800 km to 700 km takes 0.88 years and the rest from
700 km to 400 km (assuming the same thrust as at 700 km) takes 2.4
years, thus altogether 3.25 years. During solar maximum the
800$\to$700 km deorbiting takes 0.25 years and 700$\to$400 km 0.7
years, thus altogether 0.96 years. These estimates are conservative
since in reality plasma density and oxygen concentration and hence
plasma brake thrust continue to grow below 700 km.

\begin{figure}
\includegraphics[width=8.5cm]{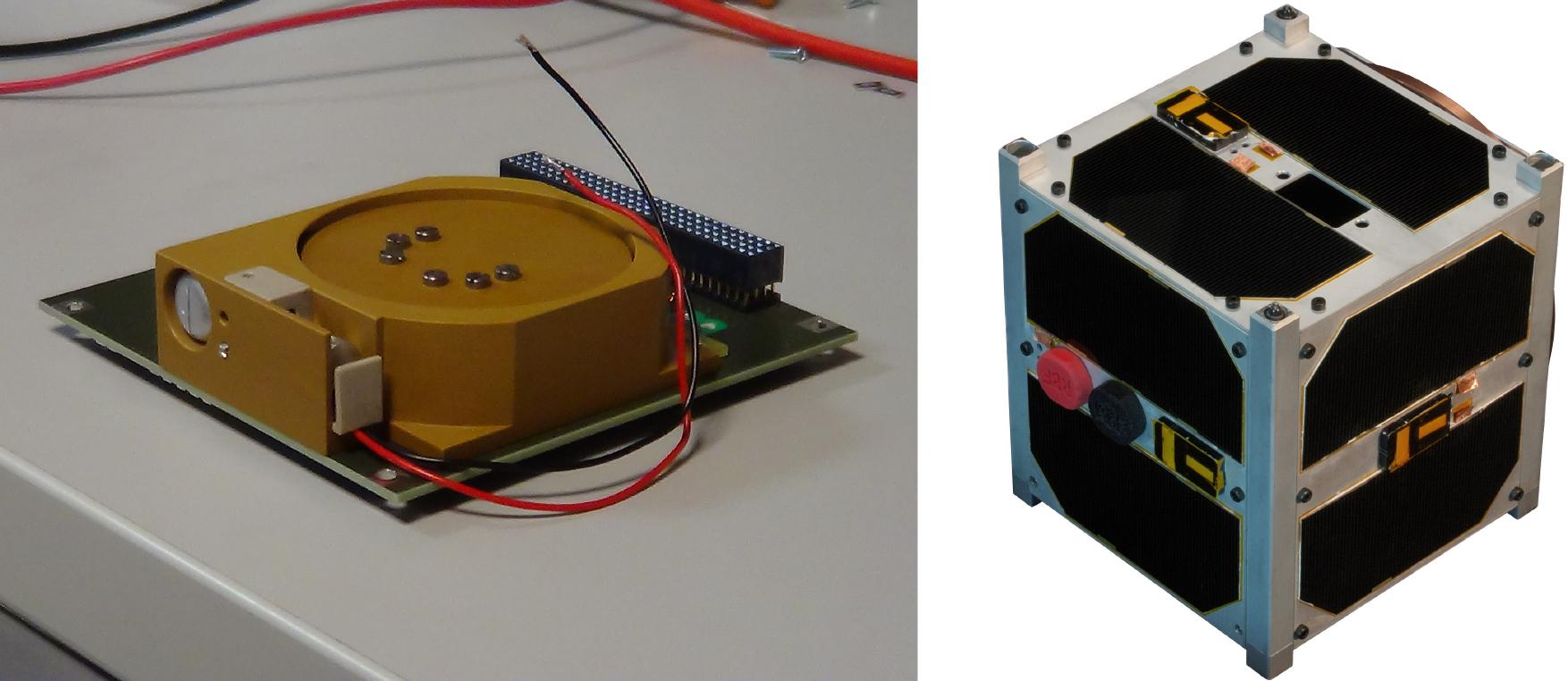}
\caption{
Left: ESTCube-1 Coulomb drag experiment payload (10x10 cm circuit board), Right:
ESTCube-1 1-U CubeSat. Photographs by Jouni Envall and Tartu Observatory.
}
\label{fig:EstcubeCard}
\end{figure}

\section{ESTCube and Aalto test missions}

The ESTCube-1 nanosatellite (1 kg, 1-U CubeSat) was launched in May 7,
2013 onboard Vega from Kourou, French Guiana. The primary payload of
ESTCube-1 is a Coulomb drag experiment \citep{EnvallEtAl2014} and the goal of the
mission is to measure the Coulomb drag force acting on its
10 m long tether in LEO conditions (670 km altitude polar orbit). The
satellite can polarise the tether to $\pm 500$ V, hence both E-sail
(positive polarity) and plasma brake (negative polarity) experiments
can be carried out. The positive mode uses nanographite electron guns
while the negative mode in this case needs only a voltage source which
creates a potential difference between the tether and the satellite
body: electron collection by the satellite's conducting parts can
balance the ion current gathered by the tether.

The ESTCube-1 tether experiment has not yet been started. The
satellite is otherwise working fine, but there are some technical
issues with its attitude control system. Work is ongoing to resolve
those issues by software and we hope to be able to carry out the
tether experiment soon. All aspects of ESTCube-1 are documented in an
ESTCube-1 special issue of the \emph{Proceedings of the Estonian
  Academy of Sciences} which is expected to appear in late May 2014.

Aalto-1 is a Finnish 3-U CubeSat which will be launched in
2015 \citep{KestilaEtAl2013}. Aalto-1 carries (among other things) a 100 m long Coulomb drag
tether experiment. The work of building the satellite and the payload
is proceeding at an advanced stage.

\section{Technology roadmap}

The E-sail and the plasma brake have a lot of technical synergy. The
baseline concept is that the interplanetary E-sail uses positive
voltages, electron gun and multiple centrifugally stabilised tethers
while the plasma brake uses vertical gravity-stabilised tether(s),
negative voltage and no electron or ion gun (a relatively small
conducting object being enough to gather the balancing electron
current in that case). However, it is not impossible to combine the
elements of the technology also in other ways.

One LEO CubeSat experiment (ESTCube-1) is already in orbit and another
one (Aalto-1) is being built. If these experiments are technically
successful, they will demonstrate deployment of 10 m and 100 m tether,
respectively, and will measure the strength of negative and positive
polarity Coulomb drag effects in LEO plasma conditions.

Our current plan for the next step is to fly a 3-U CubeSat experiment
(ESTCube-3) in solar wind intersecting orbit which measures the E-sail
effect with a 1 km long tether using 5-10 kV voltage. Because launch
opportunities to solar wind intersecting orbit (for example a lunar
orbit) are less frequent than ordinary LEO CubeSat launches, to ensure
mission success we plan to prove the satellite's technologies by first
flying an identical satellite (ESTCube-2) in LEO. ESTCube-2 also
naturally demonstrates a 1 km long plasma brake. After ESTCube-3, we
will have a measurement of the strength of the E-sail effect in the
actual environment (solar wind), a demonstration of deploying a 1 km
long tether and a demonstration of using the E-sail effect for
spacecraft propulsion.

After ESTCube-3, we need an E-sail ``pathfinder'' mission (comparable
to SMART-1 in its philosophy) which tests the use of E-sail propulsion
for going to some target and carries some payload. For example, it
could be a NEO mission equipped with imaging instruments.

\section{Discussion and conclusions}

Scaling up the manufacturing capacity of tethers and demonstrating
their robust deployment in laboratory and in space is important and
serves all applications of Coulomb drag devices. For the E-sail, a
specific challenge for raising the TRL is to get affordable launch
opportunities into the relevant environment i.e.~the solar
wind. Demonstration of multi-tether centrifugal deployment is
important as well, but it can be rather well simulated numerically and
can also be demonstrated in LEO.

In conclusion, the development of Coulomb drag devices (E-sail and
plasma brake) has gone well and their performance and other
characteristics look very promising. While the possibility of negative
surprises can never be excluded, presently we are not aware of any
unsolved issues that could make it difficult to construct efficient
plasma brake deorbiting devices and production scale interplanetary
E-sails.

\emph{Acknowledgements.} This research was financed within the
European Community's Seventh Framework Programme ([FP7/2007-2013])
under grant agreement number 262733 and the Academy of Finland grant
decision 250591.

\end{document}